\documentclass[12pt,a4paper]{article}
\usepackage{graphics,graphicx}
\usepackage{amssymb,epsfig,amsmath,euscript,array}

\makeatletter
\@addtoreset{equation}{section}
\makeatother

\makeatletter
\@addtoreset{equation}{section}
\makeatother

\newcounter{multieqs}

\newcommand{\be}{\begin{equation}}
\newcommand{\ee}{\end{equation}}

\begin{document}
\begin{flushright}
QMUL-PH-05-04\\
hep-th/0502228

\end{flushright}

\vspace{20pt}

\begin{center}

\Large\textbf{Geometrical Tachyon Kinks}
\vspace{0.2cm}
\Large\textbf{and $NS$5 Branes.}
\vspace{33pt}

\centerline{\normalsize \bf Steven Thomas\footnote{e-mail:
s.thomas@qmul.ac.uk} \ and
John Ward\footnote{e-mail:
j.ward@qmul.ac.uk} }

{\em \normalsize Department of Physics\\
Queen Mary, University of
London\\ Mile End Road\\ London, E1 4NS U.K.\\}

\vspace{40pt} {\normalsize\bf Abstract}

\end{center}

We further investigate the $NS$5 ring background using the tachyon map.
Mapping the radion fields to the rolling tachyon
helps to explain the motion of a probe $Dp$-brane in this background. 
It turns out that the radion field becomes tachyonic when
the brane is confined to one dimensional motion inside the ring.
We find explicit solutions for the geometrical tachyon field that describe
stable kink solutions which are similar to those of the open string tachyon.
Interestingly in the case of the geometric tachyon, the dynamics is 
controlled by a cosine potential. 
In addition, we couple a constant electric field to the probe-brane, but find 
that the only stable kink solutions occur when there is zero electric  
field or a critical field value. We also investigate the behaviour of Non-BPS 
branes in this background, and find that the end state of any probe brane
is that of tachyonic matter 'trapped' around the interior of the ring. 
We conclude by considering compactification 
of the ring solution in one of the transverse directions.

\newpage
\section{Introduction.}

One of the many outstanding problems in string theory concerns dynamics in time
dependent backgrounds. Fortunately there has been much recent progress made in this
area, due in part to the seminal work of Sen \cite{sen1, sen2}, which showed that 
the rolling tachyon could be used to represent the decay of an unstable $D$-brane
or brane-antibrane pair. Surprisingly the DBI action can be used to
describe much of the dynamics of this tachyon \cite{sen4, kluson4, roo, garousi}.

It is well known that in type II string theory there are BPS and Non-BPS
$Dp$-branes. The Non BPS-branes are unstable due to the presence of an open
string tachyon on their worldvolume. As this tachyon condenses, the brane can decay to form a 
new $D$-brane configuration which is stable. There has been much work on the
various solutions (see \cite{sen3, kim, kim2, brax} for
example) which show that the tachyon can have a variety of kink or vortex solutions.

A novel approach to the problem of time dependence was initiated in \cite{kutasov}, where the time
dependent motion of probe $Dp$-brane in a coincident $NS$5-brane background was
considered. In that paper Kutasov showed that there was a map between the radial field living on
the $D$-brane world volume and the rolling tachyon associated with an unstable $D$-brane. Thus the
motion of the probe brane in the throat could be described by tachyon condensation
\footnote{Extensions of this work can be found in 
\cite{sahakyan, thomas,Nakayama, Nakayama2, Chen, chen2, thomas2, bak, ghodsi, yavartanoo, panigrahi, 
peet, kluson, huang}.}.
This is possible since the $Dp$-brane breaks all the supersymmetries, and is therefore unstable. The
probe brane decays into tachyonic matter which has a pressure that goes exponentially to zero at late times.

Furthermore, in \cite{kutasov2} it was shown that by compactifying one of the
transverse directions to the source branes, it was possible to obtain a tachyon
potential which had the same functional form as that derived using string field theory
techniques. This point was further explored in \cite{kluson2, kluson3, kluson5} using
Non-BPS $Dp$-branes. The main result of this work has been to suggest that
the tachyon may have a geometrical origin, although there are still many open
questions relating to this.

In this paper we will extend the analysis begun in \cite{thomas} to consider
whether there is a map between the radion and tachyon fields on a probe $Dp$-brane
in an $NS$5-brane ring background. Since there is a compact plane inside the ring,
it will be useful to see if this also yields a suitable tachyonic potential.
We will begin with a review of the ring background, and the effective action
for the BPS $Dp$-brane. We will first try and establish if there is a map
between the radial mode and that of the rolling tachyon in the various
parts of the background, before proceeding to look for kink solutions. 
We will then consider the introduction of a Non-BPS brane inside the ring, and
investigate the dynamics when the tachyons have temporal or spatial dependence.
In the last section, we will consider the effect of compactifying one of the
directions transverse to the ring plane to see if this yields new information on
the geometrical origin of the tachyon. We close with some remarks and some
possible extensions for future work.

\section{Ring background.}
We begin with a review of the background space-time, modified by the presence
of a ring of $k$ static $NS$5-branes. As usual we write down the corresponding
CHS solution for the background \cite{chs}.
\begin{eqnarray} \label{eq:CHS}
ds^2&=& dx_{\mu} dx^{\mu} + H(x^n) dx^m dx^m\nonumber\\
e^{2(\phi-\phi_0)}&=&H(x^n)\nonumber \\
H_{mnp}&=&-\epsilon^q_{mnp} \partial_q \phi.
\end{eqnarray}

Where $\phi$ is the value of the dilaton on the probe brane and $H_{mnp}$ is the 3-form field strength for
the $B$ field. The Roman indices run over the four transverse directions, and $H(x^n)$ is the harmonic function solving the Poisson equation for the $NS$5-branes in the
transverse space. The harmonic function describing a ring in the $x_6 - x_7$ plane
is obtained by taking the extremal limit of the rotating $NS$5-brane solution \cite{sfetsos},
or by solving for the appropriate Greens function. In any case, the solution is given by
\begin{equation}\label{eq:harmon}
H=1+\frac{kl_s^2}{\sqrt{(R^2+x_6^2+x_7^2+x_8^2+x_9^2)^2-4R^2(x_6^2+x_7^2)}},
\end{equation}
where $l_S$ is the string length. This represents a continuous distribution of
fivebranes around a ring of radius
$R$ in the transverse space. In \cite{sfetsos} the exact harmonic function for $k$ 
NS5 branes around a ring was obtained. If we consider the case where $k$ is large 
then (\ref{eq:harmon}) can be thought of as an approximation to the exact harmonic function for distances 
sufficiently far from the ring so that individual NS5-branes are not resolvable.
This approximation simplifies calculations considerably and we shall adopt it throughout. It would be 
interesting to see whether results can be obtained for the exact ring harmonic function. 

 Since the coincident fivebrane background has a transverse
$SO(4)$ symmetry which is effectively broken down to $SO(2) \times SO(2)$ by the ring solution,
it will be more convenient to  use polar coordinates.
\begin{eqnarray}
x_6 &=& \rho \cos(\theta), \hspace{0.4cm} x_7 = \rho \sin(\theta)\nonumber \\
x_8 &=& \sigma cos(\phi), \hspace{0.4cm} x_9 = \sigma sin(\phi).
\end{eqnarray}
With this change of coordinates the harmonic function reduces to
\begin{equation}\label{eq:harmonic}
H(\rho, \sigma) = 1+\frac{kl_s^2}{\sqrt{(R^2+\rho^2+\sigma^2)^2-4R^2\rho^2}}.
\end{equation}
As discussed in \cite{kutasov} in order to probe this geometry it is useful to introduce
a $Dp$-brane, whose dynamical behaviour is governed by the effective DBI action.
We assume that the probe is oriented 'parallel' to the fivebranes and use the
reparameterization invariance to go to static, or Monge, gauge. Thus the action for the probe brane
becomes
\begin{equation}\label{eq:dbi}
S=-\tau_p \int d^{p+1}\zeta e^{-(\phi-\phi_0)} \sqrt{-det(G_{\mu \nu}+B_{\mu \nu}+\lambda F_{\mu \nu})}.
\end{equation}
Where $\tau_p$ is the tension of the $Dp$-brane, $F_{\mu \nu}$ is the $U(1)$ gauge
field strength, whilst $G_{\mu \nu}$ and $B_{\mu \nu}$ are the pullbacks of the
metric and B-field to the brane respectively:
\begin{equation}
G_{\mu \nu} = \partial_{\mu}X^A \partial_{\nu}X^B G_{AB}(X) \nonumber
\end{equation}
\begin{equation}
B_{\mu \nu} = \partial_{\mu}X^A \partial_{\nu}X^B B_{AB}(X).
\end{equation}
Note that $A, B = 0, 1, \ldots, 9$ run over the ten dimensional bulk space-time,
where $G_{AB}, B_{AB}$ are the bulk metric and B-field. 
As in \cite{thomas} we will neglect the contribution from the Kalb-Ramond field in the following discussion.
\section{Tachyon map.}
The time dependent dynamics of the probe brane in this particular background were discussed
in \cite{thomas} using numerical methods.
In this section we introduce the tachyon map \cite{kutasov, panigrahi} , which we hope will
shed new light on the solutions, and also give us more understanding of the
behaviour of tachyons in string theory.

Upon substitution of the background metric into our DBI action (\ref{eq:dbi}) for
time dependent scalar fields, and setting the $U(1)$ gauge field to zero 
(we will discuss non-vanishing electric fields later) we obtain
\begin{equation}
S = -\tau_p \int d^{p+1} \zeta H^{-1/2}\sqrt{1-H(\rho,\sigma)(\dot{\rho}^2 + \dot{\sigma}^2)},
\end{equation}
where we have also set the angular terms to zero to consider purely radial motion. It is important
to remember that the action is only well defined
if the higher order derivatives are vanishingly small.
This form of the action is reminiscent of that for an open string tachyon, which is
governed by a Born Infeld action of the form \cite{sen4, kluson4, roo, garousi}
\begin{equation}
S = -\int d^{p+1} \zeta V(T)\sqrt{1+\partial_{\mu}T \partial^{\mu}T },
\end{equation}
where $V(T)$ is the tachyon potential (there are also other actions describing the
behaviour of the open string tachyon see \cite{Minahan, Gerasimov} and references therein, which are
more appropriate in other regions of field space). The tachyon potential is assumed to be an even, runaway function of
$T$, with the maximum value occurring at $T=0$, and tending to zero as 
 $T \to \pm \infty$.
In fact it is possible to define a map from one action to the other, whereby we rescale
our 'radion' fields $\rho , \sigma $to become 'tachyonic' fields with a potential given by
\begin{equation}
V(T)=\frac{\tau_p}{\sqrt{H(\rho, \sigma)}}.
\end{equation}
We can consider 3 different types of motion for the probe brane
in this background, namely motion in the ring plane with $\rho < R$, motion in the
ring plane with $\rho > R$ and motion completely perpendicular to the ring plane.
We will study each of these cases separately.
\subsection{Inside the ring.}
Setting $\sigma = 0$ and assuming that $\rho < R$ we find that the harmonic
function reduces to
\begin{equation}
H(\rho) = \frac{kl_S^2}{R^2-\rho^2},
\end{equation}
where we have dropped the factor of unity since we know the probe brane will be
near the fivebranes and $\rho=\sqrt{x_6^2+x_7^2}$. Although using polar coordinates are more useful for considering
the probe brane dynamics, we will revert to Cartesian coordinates 
${x_i}, i=6, 7$. It is more
usual to consider tachyon mapping as being one dimensional. In what follows we will consider the brane to start at $x_i=-R$, and
follow its motion through the origin until it reaches $x_i=+R$.
The tachyon map in this instance is given by
\begin{equation}
T(x_i)=\int \sqrt{H(x_i)} dx_i,
\end{equation}
which can be integrated to give
\begin{equation}\label{eq:tachyon1}
T(x_i) = \sqrt{kl_s^2}\arcsin(x_i/R),
\end{equation}
and therefore
\begin{equation}
H(T) = \frac{kl_s^2}{R^2\cos^2(T/\sqrt{kl_s^2})}.
\end{equation}
We know that $\rho=0$ is an unstable point \cite{thomas}, since a probe brane initially located at
the origin will move toward
the ring if perturbed, and from the tachyon map we find that $T(x_i)=0$ at this point. The maximum values
of the field are $\pm \pi \sqrt{kl_s^2}/2$, which occur when the probe brane hits the ring.
It will be convenient to write the corresponding tachyon potential as
\begin{equation}\label{eq:potential1}
V(T)= \tau_p^{\rm unstable}\cos(T/\sqrt{kl_s^2}),
\end{equation}
where
\begin{equation}
\tau_p^{\rm unstable} = \frac{ \tau_p R}{\sqrt{kl_s^2}}.
\end{equation}
It is interesting to note that the tension of the unstable brane at this point is
proportional to the radius of the ring. The tachyon potential profile has its maximum at $T=0$, and tends
to the value $\pm \pi \sqrt{kl_s^2}/2$ as $\rho \to \pm R$, corresponding to the
point where the probe is attached to the ring. This agrees with the general descriptions of the potential proposed
in \cite{sen1, sen2} if we consider $kl_s^2 >> 1$.
The potential contains the mass of this tachyonic field, which can be seen by expanding about $T=0$, corresponding
to the perturbative vacuum. The result is
\begin{equation}
S \approx -\int d^{p+1}\zeta \tau_p^{\rm unstable} \left(1-\frac{T^2}{2kl_s^2}+ \ldots \right)\sqrt{1-\dot{T}^2},
\end{equation}
which implies that
\begin{displaymath}
M_{T}^2 = \frac{-1}{kl_s^2}.
\end{displaymath}
Note that this is small when compared to the usual (mass)$^2$ term for the open string
tachyon, $M_{T}^2=-1/2$ (in units where $\alpha' =1$) .

We can also calculate the components of the energy momentum tensor associated with
this tachyon, which we will use later.
\begin{equation}\label{eq:stress_tensor}
T_{00} = \frac{V(T)}{\sqrt{1-\dot{T}^2}} \nonumber
\end{equation}
\begin{equation}
T_{ij} = -\delta_{ij}V(T)\sqrt{1-\dot{T}^2},
\end{equation}
where the pressure goes to zero at late times as expected for tachyonic matter. This
can easily be seen since $V(T)$ tends to zero as the tachyon rolls toward its maximum,
or minimum values.
\subsection{Outside the ring.}
In this case we have $\rho > R$ and so the harmonic function becomes
\begin{equation}
H(\rho)= \frac{kl_s^2}{\rho^2-R^2}.
\end{equation}
Using the same method of analysing the tachyon map as in the previous section, with $x_i \ge R$ and $x_i \le -R$ being the
allowed range of the probe, we obtain
\begin{equation}
T(x_i)=\ln\left(\frac{|x_i|}{|R|}+\sqrt{\frac{x_i^2}{R^2}-1}\right).
\end{equation}
This tachyon is zero at the ring $x_i =\pm R$, and tends to $\infty$ as $|x_i|>> |R|$. We divide the solutions into two categories,
 namely those where $|x_i| >> |R|$, and $|x_i| \approx |R|$.
\begin{eqnarray}
T(x_i)&\to& \ln\left(\frac{|2x_i|}{|R|}\right), \hspace{0.5cm} |x_i| >> |R| \nonumber\\
T(x_i)&\to& \ln\left(\frac{|x_i|}{|R|}\right), \hspace{0.5cm} |x_i| \approx |R|,
\end{eqnarray}
and consequently the tachyon potential becomes
\begin{eqnarray}
V(|x_i| >> |R|)&=& \frac{R}{2\sqrt{kl_s^2}}\sqrt{e^{2T}-4} \nonumber\\
V(|x_i| \approx |R|)&=&\frac{R}{\sqrt{kl_s^2}}\sqrt{e^{2T}-1}.
\end{eqnarray}
This potential vanishes at $T=0$ where the probe brane hits the ring, and as
expected it gives us a pressure-less fluid at late times. The form of the potential
indicates that the probe brane will be gravitationally attracted to the ring, which
is what we would expect from \cite{kutasov}, \cite{thomas}. However, if it is 
expanded for small $T$ we find that there are positive (mass)$^2$ terms, and so we see
that the radion field cannot be redefined to be tachyonic.
\subsection{Transverse to the ring.}
If we now consider the case of motion transverse to the ring, ie with $\rho=0$ and $\sigma = \sqrt{x_8^2+x_9^2}$, the
harmonic function becomes
\begin{equation}\label{eq:transverse_harmonic}
H(\sigma) = \frac{kl_s^2}{R^2+\sigma^2}.
\end{equation}
Since we will consider motion that passes directly through the origin, we again resort to Cartesian
coordinates. Performing the tachyon map yields the following solutions as a function of $x_j, j=8,9$.
\begin{eqnarray}
T (x_j)&=&\sqrt{kl_s^2} {\rm arcsinh} (x_j / R), \nonumber \\
V(T)&=&\frac{R}{\sqrt{kl_s^2}}{\rm csgn}\left[\ {\rm cosh}(T/\sqrt{kl_s^2})\right]\ {\rm cosh}(T/\sqrt{kl_s^2}),
\end{eqnarray}
where the csgn function is defined to be
\begin{equation}
\textrm{csgn}[y]= \Big \lbrace \begin{array}{cc}\ 1 \hspace{0.5cm} Re(y) >1 \\
-1 \hspace{0.5cm} Re(y) < 1 \\ \end{array}.
\end{equation}
Thus we find that at $T=0$ the tachyon potential is at a minimum, whilst
for $T\to \pm \infty$ we have $V(T)\to \pm \infty$. Immediately this suggests
that we are not dealing with a true tachyon, since we know that the tachyonic potential has a maximum at $T=0$ and a minimum 
at $\pm \infty$. Furthermore, the field is free to oscillate about the minimum of the potential which
implies that it will be massive. However it does
explain why we find the interesting oscillatory behaviour described in \cite{thomas}, since as this 'pseudo-tachyon' moves toward 
its lowest energy configuration, it will roll around the minimum of the potential passing through the origin as it does
so. Eventually we expect that it will radiate away its energy via closed
string emission and come to rest
at the origin, which we know to be an unstable point. Thus it would appear that
the probe brane is ultimately doomed to hit the ring.

We have seen that the tachyon map is defined in each of the three cases, but
that only in the case where we have motion inside the ring do we actually
recover a 'real'(i.e. negative mass$^2$) tachyonic field. We will look at this in some more detail in the next section.
\section{Geometrical tachyon kinks.}
Using our tachyon map we are able to rotate our radion field to become tachyonic. This is
due to the fact that we are considering motion in a compact space bounded by the ring of $NS$5-branes at a radial
distance $R$ from the origin. We will call this field a geometrical tachyon, so as to avoid
confusion with the open string tachyon, and denote it by $\tilde T$. We will also write the unstable
tension as being $\tau_p^u$.
The action in terms of our new tachyon is given by
\begin{equation}\label{eq:geometrical_action}
S = -\int d^{p+1} \zeta V(\tilde T) \sqrt{-det(\eta_{\mu \nu}+
\partial_{\mu} \tilde T \partial_{\nu} \tilde T +\lambda F_{\mu \nu})},
\end{equation}
where
\begin{equation}\label{eq:cosine}
V(\tilde T) = \tau_p^u \cos\left(\frac{\tilde T}{\sqrt{kl_s^2}}\right)
\end{equation}
For simplicity we consider the case directly related to the rolling radion mode,
namely a time dependent tachyon. We will also set the gauge field to zero for the moment.
The energy momentum tensor
associated with such the tachyon field is given by  (\ref{eq:stress_tensor})
with $T$ replaced by $\tilde{T}$. We could solve the full equations of 
motion to determine the dynamical
behaviour of the tachyon field, however it is far simpler to use
the conservation of the energy momentum tensor given by
\begin{equation}
\partial_{\mu}T_{\mu \nu} = 0,
\end{equation}
which shows us that the $T_{00}$ component must be independent of time.
 This allows us to
write the first equation above as
\begin{equation}
\frac{V(\tilde T)}{\sqrt{1-\dot{\tilde T}^2}} = \gamma = \rm{constant.}
\end{equation}
Upon substitution of the potential, we can integrate this equation to determine the
time dependence of the tachyon field. As an intermediate step we write \cite{kim}
\begin{equation}
(\partial_t \tilde T)^2 + \frac{V(\tilde T)^2}{\gamma^2} = 1.
\end{equation}
This tells us immediately that there are no kink solutions\footnote{Timelike 
kinks usually correspond to S-branes \cite{sbrane1, sbrane2}, with
a Euclidean DBI action} possible for
\begin{equation}\label{eq:constraint_1}
\left(\frac{\tau_p^u}{\gamma}\right)^2 > 1.
\end{equation}
In addition, we see that if the above condition becomes an equality, then the only solution we expect
to obtain will be the trivial $\tilde{T}(t)=0$. 
Performing the integral gives us, up to any arbitrary constants,
\begin{equation}
\frac{\sin(\tilde T/ \sqrt{kl_s^2})}{\sqrt{1-u^2\cos^2(\tilde T /\sqrt{kl_s^2})}} = Sn\left [\frac{t}{\sqrt{kl_s^2}},u \right ]
\end{equation}
where we have written $u= \tau_p^u/\gamma$, and $Sn$ is the Jacobi Elliptic function. This is
actually invertible and we obtain the following solution for the tachyon.
\begin{equation}\label{eq:tachyon_soln1}
\tilde T(t) = \sqrt{kl_s^2} \arcsin \left( \frac{Sn\left [\frac{t}{\sqrt{kl_s^2}},u\right ]\sqrt{1-u^2}}{\sqrt{1-u^2Sn^2\left[\frac{t}{\sqrt{kl_s^2}},u\right]}}\right).
\end{equation}
Now, using the conservation equation we see that at $t=0, \tilde T =0$, which gives us a constraint
on the allowed values of $u$. In fact we obtain
\begin{equation}\label{eq:constraint_2}
(\partial_t \tilde T)^2 = (1-u^2),
\end{equation}
implying $u$ lies in the range $0 \le u \le 1$ in agreement with (\ref{eq:constraint_1}). We
also remind the reader that $u=1$ corresponds to $\tilde T=0$, and that if $u=0$ then
the tachyon is moving at the speed of light. 

The solution (\ref{eq:tachyon_soln1}) is effectively the equation of motion for the $Dp$-brane
discussed in \cite{thomas}. In that analysis we had to resort to numerical simulation to
determine the probe brane dynamics, however using the tachyon map has yielded an
explicit solution. We plot solutions for various values of $u$ in Figure 1. It is
interesting to see that in the $u \to 0$ limit, the motion of the probe brane is
ultra relativistic. Whilst for $u\to1$ the probe accelerates toward the ring with much
smaller velocities.
This solution is intuitively understood, since if there is tachyon rolling, the decreasing potential must be compensated
by an increase in the derivative in order for $\gamma$ to remain constant (unless, of course, the
tachyon field is moving at the speed of light). 
 Not plotted
in this figure is the $T(t)=0$ solution, which corresponds to the probe brane being trapped at
the origin.

\begin{figure}\begin{center}
\includegraphics[width=8cm,height=7.5cm]{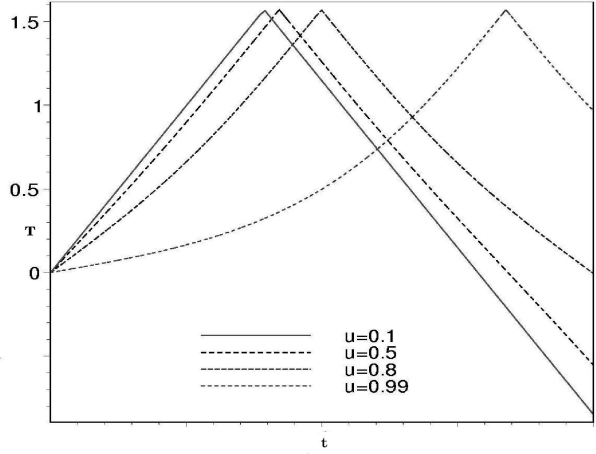}
\end{center}
\vspace{-1.5cm}\caption{Solution curves for the evolution of the time dependent tachyon
 ${\tilde T}$ with varying values of $u$. The maximum
value of the tachyon field is $\pi /2$, which corresponds to the probe brane being stuck to the ring,
 thus the continuation of the curves beyond the time this happens is unphysical. For simplicity we 
have set $\sqrt{kl_s^2} =1$. }
\end{figure}

We may expect to find a kink solution if we consider the tachyon to be dependent upon a solitary
spatial direction, namely $\tilde T=\tilde T(x)$. In this case the conservation of energy momentum tensor implies
\begin{equation}
\frac{V(\tilde T)}{\sqrt{1+(\partial_x \tilde T)^2}} = \gamma =\rm{constant.}
\end{equation}
An initial inspection of this equation reveals that there is no kink solution
if $u<1$, and that if this constraint becomes an equality then
again, the only solution will be $\tilde T(x) = 0$.
From the conservation equation we expect that $\gamma$ will be zero. This is because a kink
solution requires $\tilde T(x)=\pm \pi\sqrt{kl_S^2}/2$ for $x \ne 0$, and consequently the potential
must be zero. This implies that $\gamma$ is zero  since the derivative of $\tilde T$ will not be infinite.
At $x=0$ we find that the derivative blows up in the denominator, and so once again we find that $\gamma =0$.
It will transpire \cite{sen3} that $\gamma$ is the width of the kink, and since we expect it to be zero
this implies that a BPS $D(p-1)$ brane has zero thickness.

If we now proceed with our integration, we find that the solution is given by
\begin{equation}\label{eq:tachyon_soln2}
\tilde T(x) = \sqrt{kl_s^2} \arcsin \left( \frac{\sqrt{u^2-1}}{u} Sn \left[\frac{xu}{\sqrt{kl_s^2}},\frac{\sqrt{u^2-1}}{u} \right]\right),
\end{equation}
where once again $u=\tau_p^u/\gamma$ and $Sn$ is the Jacobi elliptic function. Now, we make an important
observation. For small $\gamma$ we find $u >> 1$, and so the second term in the Jacobi function can
be approximated by unity. Using the properties of $Sn$, namely that
\begin{displaymath}
Sn(z,1)=\tanh(z),
\end{displaymath}
our expression for the tachyon reduces to
\begin{equation}\label{eq:tachyon_soln2a}
T(x)=\sqrt{kl_s^2}\arcsin \left( \tanh\left[\frac{xu}{\sqrt{kl_s^2}}\right] \right).
\end{equation}
This is clearly a kink solution, which interpolates between $\pm \sqrt{kl_s^2}\pi/2$ due to the arcsin function,
for non-zero $x$, whilst at $x=0$ we find $T(x)=0$. Furthermore by differentiating the full solution we find that
at $x=0$ we have
\begin{equation}
(\partial_x \tilde T)|_{x=0} = \sqrt{u^2-1}
\end{equation}
This can be made infinite by sending $u$ to infinity, and so we recover the usual solution for
tachyonic kinks \cite{sen3}. In terms of the bulk picture, this corresponds to
a brane attached to the ring at $-R$ for $x<0$, and at $+R$ for $x > 0$. At $x=0$ we obtain the usual soliton
solution which stretches across the diameter of the ring.
This kink solution is interesting since the geometrical tachyon only oscillates between the two zeros of the
potential, and not the two minima. The open string tachyon also has this behaviour, but it has a runaway potential
which is effectively of zero width, whereas the geometrical tachyon potential is clearly of finite width.
This is not the case for topological defect solutions in field theory, which
tend to stretch from one minima to another in order to be stable.
Furthermore, we can compute the energy density ${\cal E} $ of the kink \cite{kutasov2,sen3,kluson3} using
\begin{equation}
{\cal E}=\int_{-\infty}^{\infty} dx V(\tilde T) \sqrt{1+(\partial_x \tilde T)^2}.
\end{equation}
In the large $u$ limit we obtain, after some algebra
\begin{eqnarray}\label{eq:energy}
{\cal E} & = &\tau_p^u \sqrt{kl_s^2}\int_{-\infty}^{\infty} dy \
 \ {\rm sech}^2(y) \nonumber \\
&=& 2\tau_p^u \sqrt{kl_s^2}\nonumber \\
&=& 2R\tau_p.
\end{eqnarray}
Clearly the energy corresponds to a kink solution which is stretched across the diameter of the ring. If we compare this
to the energy bound \cite{sen3} we find
\begin{equation}
T_{\alpha \beta}^{kink} = -\eta_{\alpha \beta} \int_{-\pi\sqrt{kl_S^2}/2}^{\pi\sqrt{kl_s^2}/2} d\tilde T \ V(\tilde T),
\end{equation}
which reduces to
\begin{equation}
T_{\alpha \beta} = -2\eta_{\alpha \beta}\tau_p^u \sqrt{kl_s^2}.
\end{equation}
Thus we can see that both integrals yield the exact same result, implying that this is the lowest energy state for
the brane.

We now investigate what happens when we couple an electric field to the kink, 
creating a charged soliton. 
For simplicity we choose a constant electric field which is perpendicular to 
$ \partial_x \tilde{T}$, so e.g. $ E_i =E \delta_{ik} $ where $x^k \neq x$
Expanding the action (\ref{eq:geometrical_action}) for small $\lambda$, and incorporating the factors of
$l_s^2$ into the field definition allows us to write the components of the energy-momentum tensor as
\begin{eqnarray}
T_{\alpha \beta}&=&-\eta_{\alpha \beta} V(\tilde T) \sqrt{1-E^2+(\partial_x \tilde T)^2} \nonumber \\
T_{x x}&=& \frac{ V(\tilde T)}{\sqrt{1-E^2+(\partial_x \tilde T)^2}}.
\end{eqnarray}
There will also be a constant conserved displacement field, which can be derived by varying the action
with respect to $\dot{A}_k$.
\begin{equation}
D = \frac{E V(\tilde T)}{\sqrt{1-E^2+(\partial_x \tilde T)^2}}.
\end{equation}
We note that using a perturbative expansion in $\lambda$ allows us to consider $E\to1$, where $E=1$ is the critical value for the field.
Using the conservation of $T_{x x}$ we can write
\begin{equation}
E = \frac{D}{\gamma}=\rm{constant.}
\end{equation}
The presence of the electric field modifies the kink solution only slightly,
find
\begin{equation}
\tilde T(x)=\sqrt{kl_s^2}{\rm arcsin} \left(\frac{\sqrt{u^2+E^2-1}}{u} Sn \left \lbrack
\frac{xu}{\sqrt{kl_s^2}}, \frac{ \sqrt{u^2+E^2-1}}{u} \right \rbrack \right).
\end{equation}
There is an interesting case where the electric field takes its critical value, as we no longer have to consider
the large $u$ limit in our solution since it reproduces (\ref{eq:tachyon_soln2a}). This generally implies that the kink solution can be non-singular.
If we allow $u=1$ in the full solution, then the tachyon is entirely dependent upon $E$. In fact it reduces to
\begin{equation}
T(x) = \sqrt{kl_s^2}\arcsin\left(E Sn\left[ \frac{x}{\sqrt{kl_s^2}},E\right]\right).
\end{equation}
Using the expansion properties of the Jacobi function we find that for  
$E$ close to unity, we have a solitary kink solution of finite width,
whilst for small $E$ we find an tiny array of small kink-antikink solutions which have period $x=2n\pi\sqrt{kl_s^2}$.
We want to know which of these solutions are stable, so we must integrate the energy momentum tensor over the $x$ direction on the
world sheet. The final result is
\begin{equation}
T_{\alpha \beta} = -2\eta_{\alpha \beta} \tau_p^u \sqrt{kl_s^2} {\rm EllipticE} 
\left(\frac{\sqrt{u^2+E^2-1}}{u} \right),
\end{equation}
where we have used the periodic properties of the Jacobi functions.
This clearly shows us that the minimum energy configuration occurs when
\begin{equation}
\frac{(u^2+E^2-1)}{u^2}=1
\end{equation}
which implies that $u \to \infty$ or $E=1$. 
The first case corresponds to the uncharged kink solution of infinitesimal width, implying that
the electric flux is diluted to the point where it is effectively zero. The second solution requires that the electric field takes its critical value,
and the resulting kink solution can be deformed to one of finite width. Furthermore all the possible kink configurations will
have the exact same energy. Thus the introduction of electric flux introduces a fixed point into the theory, since a
small electric field will find it energetically more favourable to increase to its maximum size. This tells us that
the stable kink solutions will either be uncharged, or fully charged under the $U(1)$ gauge field.

Interestingly, in the time dependent kink solution with a critical field strength, we
find that $\tilde T(t)= \sqrt{kl_s^2}\pi/2$ for all time, corresponding to the probe brane being permanently attached to the ring.
This is to be expected since flux on the brane effectively increases the 'mass', forcing the probe further into the throat
generated by the fivebranes \cite{thomas2}.

\section{Non-BPS branes.}
The existence of the unstable $Dp$-brane at the point $\rho=0, \sigma=0$ is reminiscent
of a Non-BPS brane. Thus it is useful to construct the solution for a Non-BPS brane in
this ring background. Recall that a $Dp$ Non-BPS brane is related to a BPS $D(p-1)$ brane, since
the latter is a soliton on the world-volume of the former.
The Non-BPS brane action is similar to the usual DBI governing the
behaviour of the BPS $Dp$-brane, except that it has a tachyon on its world-volume, 
and an additional tachyon potential. The action\footnote{Note that we choose this form of the action rather than the alternative \cite{kutasov3, kutasov4, kluson, kluson3},
since the form of the harmonic function makes it difficult to find space-time symmetries. It would be interesting
to look for symmetries, if any, using the alternative form of the action, and compare the results to those obtained in the following section.} is of the form
\begin{equation}
S=-\int d^{p+1} \zeta V(T)\sqrt{-det( G_{\mu \nu} + \partial_{\mu}T \partial_{\nu}T)}.
\end{equation}
The form of the action suggests that the tachyon is effectively playing the role of an extra direction in field space.
Into this action we want to substitute our $NS$5-brane background, which can
easily be shown to yield
\begin{equation}
S=-\int d^{p+1} \zeta \frac{V(T)}{\sqrt{H}}\sqrt{-det(\eta_{\mu \nu} + H\partial_{\mu}X^m \partial_{\nu} X_m +\partial_{\mu}T \partial_{\nu}T)}.
\end{equation}
Where we use static gauge, and the $X^{m}$ are the transverse scalars.
We now use the tachyon map to redefine the radion field, recalling that $\tilde T$ is the geometrical tachyon
the action becomes:
\begin{equation}\label{eq:nonbps_action}
S=-\int d^{p+1} \zeta V(T,\tilde T) \sqrt{-det(\eta_{\mu \nu} + \partial_{\mu}T \partial_{\nu}T + \partial_{\mu}\tilde T \partial_{\nu} \tilde T)}.
\end{equation}
Note that we are able to couple the two tachyon potentials together, with 
 $V(T,\tilde T) = V(T){\tilde V}(\tilde T )$, ${\tilde V}(\tilde T )$ being the potential of the 
geometric tachyon. This is already suggestive of something interesting. The
geometrical tachyon potential is that already derived in (\ref{eq:potential1}).
We will try and remain general about the form of the tachyon potential $V(T)$ by
insisting that it meets the criteria described in the previous section.
Note that this behaviour is easily satisfied by the $1/\cosh(T/\sqrt{2})$ potential \cite{kutasov4, lambert}.
In addition we note that the form of the action allows the two tachyons to decouple
from each others equation of motion. Thus we may look at the dynamics by explicitly
solving these equations, or by conservation of the energy momentum tensor.
The form of the action in (\ref{eq:nonbps_action}) suggests that we could define
a complex tachyon field given by $U=T+i\tilde T$, however for the purpose of this
note we will not do so. This is usually done when we have a $Dp$ and a $\bar{Dp}$-brane and are looking for vortex solutions.
It would of course be interesting to see what the effects of this
redefinition would mean in terms of the general behaviour of the probe motion.

We can now proceed with our analysis of solutions to the equations of motion for 
the non-BPS action. 
\subsection{One spatial direction.}
To begin with we will consider the simple case where  $T=T(x)$ and $\tilde T = \tilde T(x)$, where $x$
is an arbitrary direction on the world-volume. Here and in the remainder of the section 
we shall set $l_s^2 = 1 $.  Denoting the derivative with respect
to $x$ by a prime, we can write the action as follows.
\begin{equation}
S=-\int d^{p+1} \zeta V(T,\tilde T) \sqrt{ 1+ T^{'2} + {\tilde T}^{'2} }
\end{equation}
which allows us to calculate the associated energy momentum tensor, with components
\begin{eqnarray}
T_{\alpha \beta}&=& -\eta_{\alpha \beta} V(T, \tilde T) 
\sqrt{1+T^{'2}+{\tilde T}^{'2}} \nonumber \\
T_{x x}&=& \frac{-V(T, \tilde T)}{\sqrt{1+T^{'2}+ {\tilde T}^{'2} }}
\end{eqnarray}
Where $\alpha, \beta$ run over the $0, 2 \ldots p-1$ directions perpendicular to $x$.
We will assume that the open string tachyon has the usual kink solution, namely that
 as $x \to \pm \infty $, $T(x) \to \pm \infty$ and $V(T) \to 0$.
Using the conservation of the energy momentum tensor,
 $\partial_x T_{xx}=0$ \hspace{0.1cm} $\forall x$, we
see that this is automatically satisfied by the kink solution since the open string
tachyon component of the potential
rolls to zero as the tachyon condenses. Furthermore, this is true irrespective of
the behaviour of the geometrical tachyon. In fact it turns out that the $xx$ component
of the tensor must vanish for all $x$, not just the derivative \cite{sen3}, \cite{kim}.
In any case, this physically corresponds to the appearance of a codimension 1 brane
located at the origin of the ring. This is just the BPS $D(p-1)$ probe brane used to
probe the background \cite{thomas}.
Alternatively we may find that the geometrical tachyon condenses first. In which case
the brane will be stretched across the diameter of the ring, leaving an unstable soliton
at the origin. It would be interesting to see what happens when both fields condense
at the same time.
\subsection{Two (or more) spatial directions.}
We can now extend the analysis to consider dependence upon two (or more) spatial directions,
namely $T=T(x)$, $\tilde T = \tilde T(y_j)$, where $j=2 \ldots p+1$.
 The associated components of the energy momentum
tensor are
\begin{eqnarray}
T_{\alpha \beta}& = & -\eta_{\alpha \beta} V(T,\tilde T) \sqrt{(1+(\partial_x T)^2)(1+(\partial_{y_j} \tilde T)^2)} \nonumber \\
T_{xx}& = &- V(T, \tilde T) \frac{\sqrt{1+(\partial_{y_j} \tilde T)^2}}{\sqrt{1+(\partial_x T)^2}}
\nonumber \\
T_{y_jy_j}& = &- V(T,\tilde T)\frac{\sqrt{1+(\partial_x T)^2}}{\sqrt{1+(\partial_{y_j} \tilde T)^2}}.
\end{eqnarray}
Now we find there are two conservation conditions, $\partial_x T_{xx}=0$ and $\partial_{y_j} 
T_{y_jy_j}=0$.
 These simply state that $T_{xx}$ is independent of $x$ and $T_{y_jy_j}$ is independent of the 
$y_j$'s.
We look for the usual tachyon kink solution in the $x$ direction, which satisfies both conditions 
in the limit that $x \to \pm \infty$. At the point $x=0$ we expect that the derivative of
the tachyon field becomes infinite, which means that $T_{xx}$ vanishes. But there is still
the conservation of $T_{y_jy_j}$ to consider. In order for this to hold for all $y_j$ we must
ensure that ${\tilde V}(\tilde T) \to 0$, which means that the geometrical tachyon must also condense.
But this tachyon also has a kink solution associated with it, and so the conservation conditions
are automatically satisfied.

We know that the condensation of the open string tachyon yields a BPS $D(p-1)$ brane, but we may
well enquire about what the condensation of the geometrical tachyon correspond to? Naively we may assume that it
gives rise to a Non-BPS $D(p-2)$-brane, but this brane would be unstable and have no
tachyonic modes left which could condense and stabilise it. 

More generally, using the factorization properties of the Non-BPS brane action, we may write the brane
descent relations for both kink solutions and find the energy,
\begin{equation}
{\cal E} =\int_{-\infty}^{\infty} V(T)dT \int_{-\pi \sqrt{k}/2}^{\pi\sqrt{k}/2}
{\tilde V}(\tilde T) d \tilde T.
\end{equation}
In order to do this we first specify the form of the tachyon potential, which we will take to be
\begin{equation}\label{eq:tachyon_potential}
V(T)=\frac{\tau_p^{non}}{\cosh(\frac{T}{\sqrt{2}})}.
\end{equation}
with $\tau_p^{non} $ the non-BPS brane tension.
The first integration yields
\begin{equation}
{\cal E} = \pi \sqrt{2} \tau_p^{non} \int_{-\pi \sqrt{k}/2}^{\pi \sqrt{k}/2}
{\tilde V}(\tilde T) d \tilde T  
\end{equation}
However, we have already seen that the configuration of the BPS D(p-1) brane is itself unstable 
due to the geometrical tachyonic mode also forming a kink. 
If we integrate over this potential we find
\begin{equation}
{\cal E}= 2\pi R \tau_p^{non}\sqrt{2}
\end{equation}
This can be written as
\begin{equation}\label{eq:tension}
{\cal E} = (2 R \tau_p) \times 2\pi
\end{equation}
where we have used the relation $\tau_p^{non} = \sqrt{2}\tau_p$,  $\tau_p$
being the BPS Dp brane tension \cite{kluson2}. One possible interpretation of the
form of the energy (\ref{eq:tension})is as follows. The first factor in (\ref{eq:tension})
is the energy of Dp brane stretched along the diameter of the ring, which we calculated earlier 
(\ref{eq:energy}). The additional factor of $2 \pi$ can be thought of as coming from 'smearing' the 
stretched brane around the inside of the ring.

\subsection{Space and time dependence.}
We assume that the geometrical tachyon is time dependent, whilst the open string tachyon is
spatially dependent. This allows us to write the the components of the energy momentum tensor
as
\begin{eqnarray}
T_{00} &=& V(T,\tilde T)\frac{\sqrt{1+(\partial_x T)^2}}{\sqrt{1-(\partial_0 \tilde T)^2}} \nonumber \\
T_{ij}&=& -\delta_{ij} V(T,\tilde T) \sqrt{(1+(\partial_x T)^2)(1-(\partial_0 \tilde T)^2)} \nonumber \\
T_{xx}&=& -V(T,\tilde T) \frac{\sqrt{1-(\partial_0 \tilde T)^2}}{\sqrt{1+(\partial_x T)^2}}.
\end{eqnarray}
We again appeal to the conservation equations to determine the behaviour of the kink solution.
If we assume there is a kink in the $x$ direction, then we find that there are only two possibilities
for the geometrical tachyon. We either have a kink in the time direction, or the field must
condense. Since we have already established that there is no stable kink solution, we again
find that the tachyon condenses, and this implies that the $Dp$-brane moves toward the ring.
Looking explicitly at the conservation equation for the $T_{xx}$ component we find the
expected decoupling behaviour
\begin{equation}
\partial_x \left(\frac{V(T)}{\sqrt{1+(\partial_x T)^2}}\right)=0.
\end{equation}
This allows us to integrate the equation to determine the $x$ dependence
of the open string tachyon, provided we specify
the explicit form of the potential.
If we choose the usual form (\ref{eq:tachyon_potential})
then upon integration we obtain the solution
\begin{equation}
\sinh\left(\frac{T}{\sqrt{2}}\right)=\frac{\sqrt{1-p^2}}{p}\sin\left(\frac{x}{\sqrt{2}}\right)
\end{equation}
where $p$ is an arbitrary constant of integration. If we now substitute for $\partial_x T$
in the $T_{00}$ component of the stress tensor we find
\begin{equation}
T_{00} = \frac{{\tilde V}(\tilde T)}{\sqrt{1-(\partial_0 \tilde T)^2)}}
\frac{p}{p^2+(1-p)^2 {\rm sin}^2 \left(\frac{x}{\sqrt{2}}\right)}.
\end{equation}
This is the same result that Kluson derived for branes moving on a transverse
$\mathbf R^3 \times \mathbf S^1$, and can be interpreted as an array of
$D(p-1)$-branes and $D(p-1)$-antibranes. Since there is a map between the
rolling of the time dependent geometrical tachyon and the motion of a
probe $Dp$-brane we can see that these branes simply move toward the ring.

\section{Compactification in a transverse direction.}
In \cite{kutasov2} Kutasov established the relationship between BPS and Non-BPS branes by
compactifying one of the transverse directions in the coincident $NS$5-brane background.
Since we have already seen that geometric tachyons can exist when the brane is
probing a compact space, we may well enquire if there will be tachyonic modes if we
compactify one of the transverse directions in the ring background.

Due to the symmetry of the transverse space, it is easiest to consider a compactification
in the $\sigma$ plane. We remind the reader of the harmonic function in this instance
(\ref{eq:transverse_harmonic})
\begin{equation}
H(\sigma)= \frac{kl_s^2}{R^2+\sigma^2}.
\end{equation}
We will choose to compactify the $x_8$ direction into a circle of radius $L$. The
resultant expression for the harmonic function becomes
\begin{equation}
H=\sum_{m=-\infty}^{\infty} \frac{kl_s^2}{(R^2+x_9^2)+(x_8-2 \pi L m)^2}.
\end{equation}
This sum is easy to do, since it is very similar in form to the one in \cite{kutasov2}, and
we obtain the final form of the function
\begin{equation}
H=\frac{kl_s^2}{2 L |z|}\frac{\sinh(|z|/L)}{(\cosh(|z|/L)-\cos(y/L)),}
\end{equation}
where we have defined $z=\sqrt{R^2+x_9^2}$ and $y=x_8$. In this form the harmonic function
is exactly the same as that for a coincident fivebrane background, except that if we set the
$x_9$ fields to their minimum value we find $z=R$.

Performing the tachyon map integration we obtain the following solution for the tachyon
\begin{equation}\label{eq:tachyon_equation}
\tilde T(y)=\sqrt{ \frac{2Lkl_s^2 \sinh(|z|/L)}{|z|(1+a)}}\rm{EllipticF}(\delta, r)
\end{equation}
where EllipticF is the incomplete elliptic integral of the second kind and we have made the following definitions:
\begin{eqnarray}
a &=& \cosh(|z|/L) \nonumber \\
\delta &=& \arcsin \left(\sqrt{ \frac{(1+a)(1-\cos(y/L))}{2(a-\cos(y/L))}} \right) \\
r &=& \sqrt{ \frac{2}{1+a}} \nonumber.
\end{eqnarray}
By setting the $x_9$ fields to zero we see that the behaviour of the tachyon is dependent
upon the ratios $R/L$ and $y/L$. We can see that as $y \to 0$, the numerator of $\delta \to 0$
and using the properties of elliptic integrals we find that $\tilde T \to 0$. This is to be expected
since \cite{kutasov2} essentially argues the same thing, namely that the tachyon starts at some initial
value $\tilde T = \tilde T_{max}$ at the point $y=\pi L$, and then rolls toward zero as $y \to 0$.
However we see that there is also the ratio $R/L$ in the tachyon solution, which we would expect
to explicitly determine the value of $\tilde T$ since this term dominates the cosine term associated with
the motion around the compact dimension. We can calculate the maximum value for the tachyon
\begin{equation}
\tilde T_{max} = \sqrt{\frac{2Lkl_s^2\sinh(R/L)}{R(1+\cosh(R/L))}} {\rm EllipticK} 
\left(\sqrt{ \frac{2}{1+{\rm cosh}(R/L)}} \right),
\end{equation}
where we have introduced the elliptic integral of the first kind.
We can make some approximations to determine the behaviour of the field. Firstly we take the limit $R/L << 1$
\begin{equation}
\tilde T_{max} \approx \sqrt{\frac{2Lkl_s^2 \sinh(R/L)}{R(1+\cosh(R/L))}}\ln \left( 4\sqrt{\frac{1+\cosh(R/L)}{\cosh(R/L)-1}}\right)
\end{equation}
which can be seen to tend to infinity. In the converse limit we can approximate the tachyon field by
\begin{equation}
\tilde T \approx \sqrt{\frac{2Lkl_s^2}{R}}\arcsin \left( \sqrt{ \frac{a(1-{\rm cos}(y/L))}{2(a-{\rm cos}(y/L))}}\right).
\end{equation}
which yields a maximum value of
\begin{equation}
\tilde T_{max} \approx \sqrt{\frac{2Lkl_s^2}{R}}\frac{\pi}{2},
\end{equation}
and will roll to zero as $y \to 0$. Clearly the maximum value of the tachyon field will be determined by
the exact ratio of $R/L$ and also the number of source branes.

We can also determine the tachyon potential in this instance by inverting the solution
(\ref{eq:tachyon_equation}). After some manipulation using elliptic functions we obtain
\begin{equation}\label{eq:tachyon_potential2}
\tilde V(\tilde T) = \sqrt{\frac{2LR}{kl_s^2 \sinh(R/L)}}\sqrt{\frac{(1-a)(1+a)}{1-a-2Sn^2[\Delta(\tilde T),r]}}.
\end{equation}
where $a$ and $r$ are the same as before, whilst $\Delta(\tilde T)$ is defined to be 
\begin{displaymath}
\Delta(\tilde T) = {\tilde T}\sqrt{\frac{R(1+a)}{kl_s^2L \sinh(R/L)}}.
\end{displaymath}
This is a complicated potential, and does not yield simple analytic solutions.
Furthermore we see that it is not defined for $R/L<<1$ or even $R \approx L$ due to the presence of 
the Jacobi function in the denominator. Thus, it is only valid for the $R/L >> 1$ case and we must also
assume that the tachyon never becomes too large!
The potential can only be zero if $L=0$ which means that the compact dimension is of zero size,
and a probe brane moving along it will be essentially stuck at the origin. 
For all other values of $R$ and $L$ the minimum of the potential is at some fixed non-zero value.
The unstable maximum of the potential should occur when $\tilde T= \tilde T_{max}$. For the case
when $R/L >>1$ we find that 
\begin{equation}
V_{max}(\tilde T) \approx \sqrt{ \frac{2LR \cosh(R/L)}{kl_s^2\cosh(R/L)+2\tilde T^2 R}}.
\end{equation}
As the tachyon field decreases, the potential decreases, passing through its minimum at $\tilde T=0$ 
which in the brane picture corresponds to the probe passing through the origin. Thus we 
anticipate that the tachyon field in this instance will be massive, as it was for the
case in an earlier section. This again suggests that we will obtain massive fields
when we compactify, unless the compact space is bounded by fivebranes.

Compactifying one of the directions in the plane of the ring is also possible,
however we will not consider it here. The main difficulty lies in the fact that
there is a crossover between harmonic functions in different regions of
the covering space, but there is also the additional problem of
the complicated form of the functions.
It appears likely that there will be a geometrical tachyon
in this instance when the probe is confined to the region $y=R \ldots 2 \pi L-R$, since
it is bounded by $NS$5 branes. But there is also the possibility of new geometrical tachyons
in the region $2 \pi L-R \ldots R$ which should map onto the tachyon field discussed earlier.
Once again we would expect the ratio $R/L$ to fully specify the tachyon dynamics.

\section{Discussion.}
We have investigated the dynamics of a BPS $Dp$-brane in the background of a ring of
coincident $NS$5-branes from the perspective of a rolling tachyon on the brane world
volume. We have seen that the radial mode associated with brane motion inside the ring
gives rise to a 'geometrical tachyon' on the brane, complete with a tachyon potential
which describes the changing tension.

Although we do not find a kink solution in the time-like direction, we can
find exact solutions for the rolling behaviour which are equivalent to
the equations of motion for the probe $Dp$-brane. Thus mapping the action
to a different form can simplify the problem.

We do find a spatial kink solution which appears to
be stable on its own. It still remains to be seen whether there are 
any stable vortex solutions. Additionally if one of the dimensions of the probe brane 
is wrapped on a circle then other kinds of non-singular kink solutions are possible. 
It may be interesting to investigate this and compare the results with those for the 
open string tachyon.
If the kink is charged, we would anticipate there to be a wide variety of solutions
such as kink-antikink arrays.
Furthermore we have probed the background geometry with a Non-BPS brane, which has
an open string tachyon on its world-volume. As
expected, the kink solution describes the existence of a BPS $D(p-1)$-brane but
the condensation of the geometrical tachyon implies that this brane
is somehow smeared around the interior of the ring. 
 
We have seen that there are no geometrical tachyons if we compactify one of the
transverse directions to the ring in the limit of $R/L>>1$. This is because
the compact space is not bounded by the fivebranes.
In the converse limit $R/L <<1$ we find the theory breaks down because the 'tachyonic field'
becomes infinite.
If we were to compactify along the plane of the ring, then we would expect to see two tachyonic modes, one of which should
map to the one discussed in this paper.

This analysis raises several questions about the nature of the open string tachyon in
string theory, and we can make several useful observations.
Firstly we have found that a compact space is not enough
to create geometrical tachyons. The space must be bound at either end by
fivebranes, implying that the radion field is bound between a range of values.
In \cite{kutasov2, kluson2} this corresponds to
branes located at $y=0$ and $y=2 \pi R$ on a compact circle whilst in the
ring case it corresponds to branes at $-R$ and $R$ at opposite points of the circle.
Secondly we must ensure that the compact space is symmetric about the unstable point
of the potential, to ensure that the tachyon can condense. This is equivalent to
saying that the potential must be an even function.
Thirdly it appears that geometrical tachyons can give rise to stable kink solutions on
their own, but not when we include another tachyonic field. The form of the
potential is dependent solely upon the background spacetime geometry. It would
be interesting to consider other background geometries and find different tachyonic potentials. 
This could be especially useful for cosmology, since hybrid inflation requires
the condensation of two scalar fields.


It can also be shown that if we S-dualize the background solution to obtain a ring of $D5$-branes, we
obtain similar geometrical tachyonic modes for a probe $Dp$-brane with $p=3$ and $p=5$. 
The $p=3$ case is the exact S-dual of the configuration in this note, however the $p=5$ case has
a tachyon potential given by $V(\tilde T) = \tau_p^{u} (\cos^2(\tilde T/ \sqrt{k}))$
where $k$ is a constant given in terms of the string length, string coupling and
the number of branes. This is not a truncated tachyon potential, and so we may find
stable kink solutions even when coupled to open string tachyons. It would be useful to analyse this 
and the corresponding S-dual solution for $\mathbf R^3 \times \mathbf S^1$ in the coincident background to see
if there are kink solutions, and what their implications are for the rolling tachyon. 

Finally, the emergence of a cosine potential for the geometric tachyon inside the ring
might have interesting applications to inflationary cosmology. Similar potentials have been argued to 
lead to so called natural-inflation \cite{Freese}. In our case the scales $\sqrt{k l_s^2}$ and 
$R$ of geometric tachyon potential (\ref{eq:cosine}) might play similar roles to $\Lambda$ and  $f$ 
in \cite{Freese}.However given the non-linearities of the DBI action a direct 
comparison of the tachyon potential and the cosine inflaton potentials is delicate. Rather, an analysis 
of the tachyon cosmology following e.g.\cite{Gibbons} and \cite{Li} should be carried out.   

\hspace{5.5cm} \textbf{Acknowledgements.} 

Many thanks go to Costis Papageorgakis for participating in several useful conversations
and to Konstadinos Sfetsos for useful comments.


\end{document}